\documentclass[conference]{IEEEtran}
\IEEEoverridecommandlockouts
\usepackage{cite}
\usepackage{amsmath,amssymb,amsfonts}
\usepackage{algorithmic}
\usepackage{graphicx}
\usepackage{textcomp}
\usepackage{xcolor}
\def\BibTeX{{\rm B\kern-.05em{\sc i\kern-.025em b}\kern-.08em
    T\kern-.1667em\lower.7ex\hbox{E}\kern-.125emX}}

\usepackage{tikz}

\usepackage{filecontents}

\usepackage{multirow}
\usepackage{enumitem}
\usepackage{caption}
\usepackage{multirow}
\usepackage{xurl}

\newcommand{\review}[1]{{\color{black}{#1}}}
\newcommand{\rev}[1]{{\color{black}{#1}}}

\newcommand{\roaming}{{RA}}
\newcommand{\platform}{{PA}}

\def\rot{\rotatebox}

\def\version{full} 
\def\sel{full}

\begin{document}
\addtolength{\textfloatsep}{-0.2in}
\addtolength{\dbltextfloatsep}{-0.2in}
\date{}

\title{\Large 
\bf Challenges with Passwordless FIDO2 in an Enterprise Setting: A Usability Study}

\def\plainauthor{Author name(s) for PDF metadata. Don't forget to anonymize for submission!}


\author{
    \IEEEauthorblockN{Michal Kepkowski\IEEEauthorrefmark{1}, Maciej Machulak, Ian Wood\IEEEauthorrefmark{1} and Dali Kaafar\IEEEauthorrefmark{1}}
    \IEEEauthorblockA{\IEEEauthorrefmark{1}Macquarie University}
}


\maketitle

\begin{abstract}

Fast Identity Online 2 (FIDO2), a modern authentication protocol, is gaining popularity as a default strong authentication mechanism. It has been recognized as a leading candidate to overcome limitations (e.g., it is phishing resistant) of existing authentication solutions. 
However, the task of deprecating weak methods such as password-based authentication is not trivial and requires a comprehensive approach. 
While security, privacy, and end-user usability of FIDO2 have been addressed in both academic and industry literature, the difficulties associated with its integration with production environments, such as solution completeness or edge-case support, have received little attention. 
In particular, complex environments such as enterprise identity management pose unique challenges for any authentication system. 
In this paper,  we identify challenging enterprise identity lifecycle use cases (e.g., remote workforce and legacy systems) by conducting a usability study, in which 118 professionals shared their perception of challenges to FIDO2 integration from their hands-on field experience. 
Our analysis of the user study results revealed serious gaps such as account recovery (selected by over 60\% of our respondents), and identify priority development areas for the FIDO2 community.     


\end{abstract}

\section{Introduction}

Online Authentication is critical for the security posture of every data-driven organization and is one of the main pillars of an emerging security design paradigm called zero trust~\cite{NIST_zero_trust}. Password-based methods offer limited security guarantees and Multi-Factor Authentication (MFA) techniques have been introduced as an alternative. Perhaps at the expense of usability, organizations have been slowly rolling out additional factors (e.g., in 2022, Microsoft reported 7\% increase of accounts with MFA~\cite{microsoft_report}). A \textit{something you have} factor (e.g., SMS one-time passwords) is currently a popular choice for the MFA implementation. However, as demonstrated by the recent data breaches (e.g., Uber~\cite{forbes_uber_hack}, Dropbox~\cite{dropbox_hack}, Cisco~\cite{cisco_hack}), some MFA methods continue to be vulnerable to attacks such as MFA fatigue~\cite{mfa_fatigue} or Adversary-in-the-Middle (AitM)~\cite{microsoft_aitm}.

A potential solution to address modern attack vectors on authentication, called FIDO2, has been proposed by identity experts led by the FIDO Alliance. The FIDO2 protocol was designed to provide local verification (e.g., using fingerprint or PIN), making it MFA on its own. According to digital identity hype cycle modelling by Gartner\cite{gartner_hype_cycle}, FIDO2 is expected to become a dominant solution for strong authentication in the next 2-5 years. For this to happen, FIDO2 needs to be widely adopted by the industry. However, compared to popular MFA methods, FIDO2 is a complex protocol and its security is largely dependent on its reliable implementation, deployment, and maintenance by all parties (i.e., authenticator, client, and server). 
New technologies, including FIDO2, have to overcome adaptation challenges before reaching a critical mass. While the FIDO2 security and privacy properties as well as end-user usability are well studied\cite{10.1007/978-3-030-84252-9_5,cryptoeprint:2022/084,9152694,274419}, protocol adaptation in complex environments is rarely discussed, even though technology uncertainty as defined by Stock et al. (e.g., complexity) has a major impact on integration success~\cite{Stock2004}. 

In this paper, we explore how FIDO2 as a passwordless solution and its deployability are perceived in enterprise settings. We analyze and discuss the views and experiences of \textbf{118 professionals} involved in the FIDO2 deployment and draw conclusions about challenging aspects of FIDO2 adaptability and usability, some of which can pose a serious risk to the FIDO2 popularization process.










In particular, we aimed at answering the following question:
\begin{itemize}
\item \review{What are the technological (e.g., implemented functionalities) and non-technological (e.g., know-how) challenges that discourage enterprises from integrating FIDO2-based passwordless authentication?}

\end{itemize}
To the best of our knowledge, this is the first analysis of passwordless FIDO2 in the context of practical integration with the enterprise identity lifecycle\rev{, clearly identifying challenging aspects such as account recovery, access to remote servers and missing integration guidelines. }


\section{Background}

\subsection{FIDO2} \label{sec:fido}
FIDO2 is a modern open-sourced authentication protocol, strongly supported by the major technology vendors (e.g., Microsoft, Google, Apple), and thus can be considered as a de facto standard for modern strong authentication. FIDO2 can be used as an additional factor (e.g., following a password) or as a completely independent mechanism. In the latter case, FIDO2 changes the authentication paradigm by removing the \textit{something you know} factor (usually a password, hence achieving passwordless authentication). 

\begin{figure}[t!]
    \centering
    \includegraphics[width=\linewidth]{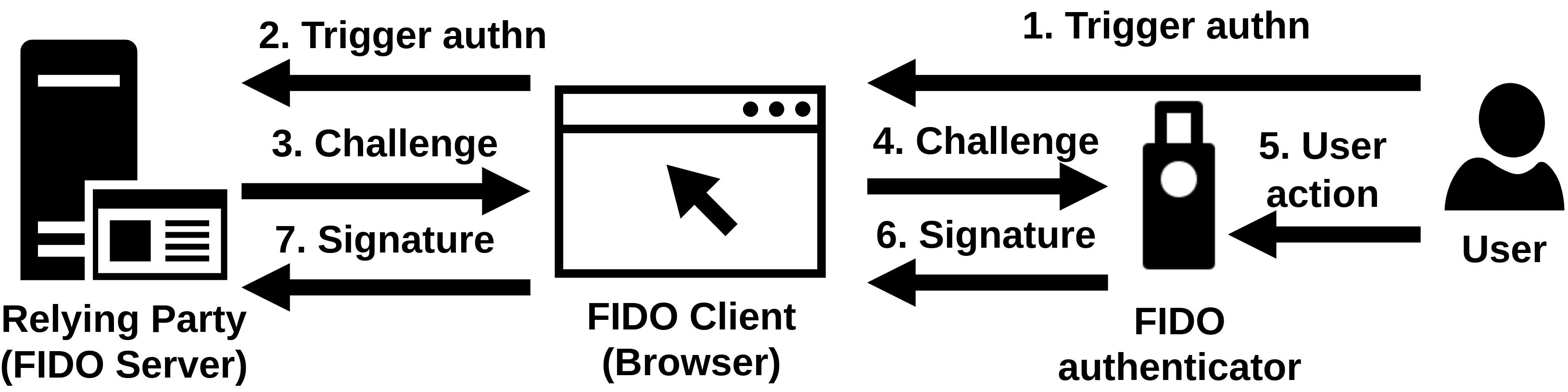}
    \caption{Overview of FIDO2 parties involved in the simplified authentication process. }
    \label{fig:simple_flow}
\end{figure}

The FIDO2 protocol involves 3 parties: FIDO server (usually integrated with Relying Party), FIDO Client (e.g., a web browser), and authenticator (e.g., USB hardware token). The communication between the parties is carried via 2 protocols: WebAuthn\cite{WebAuthnspec} and CTAP\cite{CTAPspec}. WebAuthn defines the API between client and server, whereas CTAP specifies communication to the authenticator itself. 
The user can execute two FIDO2 ceremonies: registration and authentication. The first is used to generate keys and bind them to the user's identity, and the second allows verification of key possession. The protocol flow is based on a simple challenge-response transaction (Figure \ref{fig:simple_flow}). 
The flow starts with the user requesting authentication or registration (step 1. and step 2.). The FIDO server generates a challenge and sends it to the authenticator (steps 3. and 4.). The authenticator requests a user presence or verification check (step 5.) to unlock the authenticator. In case of registration, a key pair is generated and the public key is sent back to the server. For both ceremonies, a signature is created and sent back to the server (steps 6. and 7.). The flow ends with the server verifying the signature. 

FIDO2 security strongly depends on the type (platform or roaming) and quality of authenticators. A platform authenticator (PA) is incorporated into a device (e.g., a function of Android or Windows), whereas a roaming authenticator (RA) is a separate device which can be easily detached (e.g., a USB hardware token). Authenticator vendors are mandated to use commercially recognized cryptographic primitives (e.g., ECDSA), however, the security properties of their products vary (e.g., resistance to tampering). 
The main functionality of an authenticator is to securely store private keys and execute cryptographic operations (e.g., signature). Modern authenticators use a Secure Environment or Element, and thus can achieve the highest security certifications (e.g., FIPS 140-2 or FIDO Alliance\cite{FIDOcertification}). FIDO2 is designed to mitigate a range of modern cyber-attacks. In particular, it is resistant to data breaches as servers store only public keys, and to phishing combined with AitM   because the domain is verified for each authentication. Notably, these two attacks were identified as the top cyber crimes in the Microsoft's 2022 Report \cite{microsoft_report}. 

FIDO2 represents a paradigm shift for commonly used authentication. Before FIDO2, passwordless authentication was used primarily in highly secure environments (e.g., smart cards used in government agencies and financial institutions)~\cite{doi:10.1080/1043859042000304070}. With FIDO2, the identity community pushes towards a complete migration from the \textit{something you know} into the possession factor. This drastic change raises questions about adaptation and usability. 
Lyastani et al. \cite{9152694} conducted a comparative user study on 94 participants and found out that even though FIDO2 passwordless authentication was well received, usability concerns such as recovery from the lost authenticator were present. Similarly, Owens et al. \cite{274419} examined the user's perception of passwordless authentication using \roaming s and reported users' concerns regarding availability, account recovery/backup, and setup difficulties. 

\subsection{Enterprise Setting}
The success of new technology such as FIDO2 depends not only on the end-user experience but also on how adaptation to the technology and its operation are perceived. In particular, in enterprise settings, these aspects are of pivotal importance in technology selection. 
Interestingly, this aspect of FIDO2 usability has received little attention in academia. 

The Identity and Access Management (IAM) capability in large enterprises differs from IAM in smaller organizations~\cite{doi:10.1080/23738871.2017.1398265}. The disparity can be attributed to a number of factors such as the size, complexity, use cases, technology that is used, and regulatory or legal obligations, among others. Unlike smaller organizations, large enterprises typically rely on multiple 
authentication systems to ensure the security of their assets. Moreover, these organizations are often complex and present in numerous locations around the globe. As such, authentication needs to be suitable for various devices and systems, including those 
used to access company assets. Additionally, it must also be able to cater to a large number of users, including those with various persona types that have different and often conflicting authentication requirements. 

The diversity of IAM requirements results in the growth of complexity and cost of moving from one authentication mechanism to another~\cite{10.1007/1-4020-3675-2_4}. For instance, simply `switching on' FIDO2 on a dedicated Identity Provider (IDP) is insufficient. A new mechanism is needed to address all possible authentication routes.
In such cases, enterprises must consider the types of authenticators suitable for their persona types, whether platform or roaming (hardware or software), and their security properties. They must also ensure compatibility with existing or planned authentication systems such as IDP compatibility. Additionally, some use cases may require authenticators to comply with such certifications as FIPS 140-2 or NIST 800-63B while also providing biometric-based user verification on top of simple user presence checks.

Support and toolings are equally important to allow for full authenticator life-cycle management,  including strong credential binding but also secure recovery and fallback processes. In more sophisticated cases, organizations may look into attestation to have central and policy-based control over authenticators. They may even consider baking their own, unique key into the authenticators for their workforce. 
In contrast, consumer-oriented systems prioritize usability and most FIDO2-compliant devices can be considered suitable for authentication.

\section{FIDO2 Enterprise Use Cases}

\begin{figure}[t!]
    \centering
    \includegraphics[width=\linewidth]{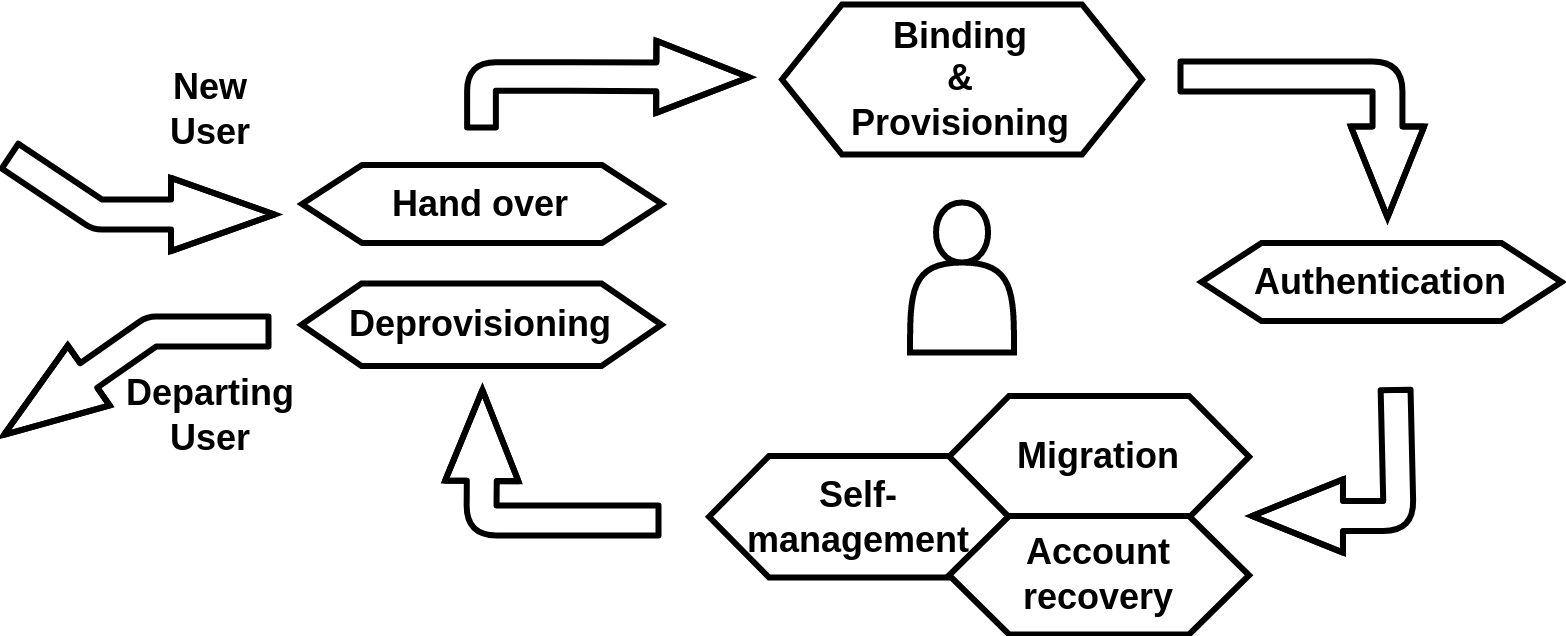}
    \caption{\small{Authentication related processes in the identity lifecycle. }}
    \label{fig:lifecycle}
\end{figure}

\review{Evaluation of security protocols is usually done in theoretical or laboratory conditions\cite{10.1007/978-3-030-84252-9_5,cryptoeprint:2022/084}, which can make it detached from real-world issues. In particular, authentication protocols are closely coupled with human-focused processes. For example, the JML (Joiner, Mover, Leaver) process is usually used in enterprises to define rules and requirements for the identity lifecycle, and by extension for authentication-related processes  (Figure~\ref{fig:lifecycle}). To understand the challenges identified in the usability study (Section \ref{sec:usabilit_study}), we provide below a brief review of the FIDO2 ecosystem  in the context of enterprise identity lifecycle use cases. 
Please note that custom or edge cases are not included, however, we believe that the described use cases, defined based on the industry-focused literature review\cite{NIST_lifecycle, SP80046r2, 7359420, FIDOAlliance_lifecycle}, give a representative set of enterprise use cases.
In particular, we discuss processes, which were found challenging by the participants of our usability study (see Section \ref{sec:usabilit_study}).
}
\subsection{Authenticator Hand Over and Binding}
The handover process is the first step to onboard a user with FIDO2 authentication. 
In case of hardware \roaming, a physical device (e.g. a USB token) needs to be provided to the user. Clearly, this operation opens a path for malicious actors to break authenticator security even before it is used. Attacks such as device cloning, firmware modifications, or key extraction in the supply chain can significantly decrease the trust for token-based passwordless schemes. 
Even though the mitigation of handover risks is out of the scope of FIDO2, vendors as well as the protocol itself provide tools to reduce some of these risks. 

Authenticator authenticity protections increase trust that hardware \roaming{} has not been tampered with. Following the guidelines provided by Pfeffer et al. \cite{272198},  software, hardware, and packaging countermeasures can be used. For software authenticity, manufacturers can execute local or remote validation of firmware based on hash or signature. Another approach is to delay the moment of firmware load and initialization till after \roaming{} is delivered. In terms of hardware authenticity, an RA  usually leverages a secure CPU (or co-processor) to perform cryptographic operations and keep the keys secure. Additionally, some vendors enhance their products with tamper-detection circuits and single-piece casts. Finally, the packaging of the delivered token can be designed to prevent tampering with the \roaming{} without visibly damaging the package or holographic sticker. 

FIDO2 was designed to provide a secure method of authentication with privacy by design to satisfy not only enterprise but also public use cases. One of the privacy-preserving mechanisms, preventing tracing of an authenticator, can be found in device attestation. The same identifier and vendor certificate is shared by a significant number of devices (over 10000~\cite{FIDOmetadata}), thus preventing unique identification. However, this feature also makes device filtering, which is a desirable security control in an enterprise, impossible. Therefore, in the recent CTAP 2.1 version~\cite{CTAPspec}, an additional way to attest a device, called ``Enterprise Attestation'', was provided. Enterprise attestation allows FIDO-relying parties to request a uniquely identifying attestation during credential registration. The details of creating authenticators with enterprise attestation are out of the scope of FIDO2 but ideas on how to do it have already been proposed by vendors (e.g., unique identifier, certificate, or entire PKI stack).

Once a user receives an authenticator, the binding process is used to create a link between (user) identity and authenticator (a public key in the case of FIDO2) \cite{SP80063B}. Interestingly, the protocol does not specify the process of verifying the identity. However, because the security of the scheme heavily depends on the proper identity assignment, we explore possible approaches to secure the binding of authenticators in an enterprise setting.

Firstly, an authenticator can be bound to the identity  
before it is delivered to the user. Such a process, either remote (e.g., \roaming{} sent to the user) or in-person is costly and difficult to scale. Alternatively, binding can be established during the first enrollment (e.g., for new employees). As described in the FIDO Alliance guidelines \cite{FIDOAllliance_best}, based on the required assurance level, one of the following models can be used. Firstly, the Trust on First Use (TOFU) model binds an authenticator to a new unknown account, the Invitation model leverages the user's pre-collected data (e.g., email address) to provide a unique binding mean (e.g., one-time use link). Additionally, a 3rd party process can be used to validate identity. For example, an external Identity Provider (Federation model) or Identity Verifier (Identity Proofing model) can certify the user's identity. Finally, authenticator binding can be required for existing users (Post-enrollment binding). In this case, an existing authentication with the equivalent security posture can be used (Anchor Model). In all cases, it is pivotal that the binding is done during a strongly established session.

\review{
The diversity and complexity of potential enrollment procedures (e.g., handover, binding, and provisioning) described here pose significant difficulties in constructing and implementing a secure sequence. It is worth noting that experts involved in our study also recognized the enrollment processes as challenging (see Section \ref{sec:usabilit_study}).
}

\subsection{Authenticator Migration}
\label{sec:auth_migration}
In an enterprise setting, devices may have a predefined lifespan (2-3 years according to the Gartner report\cite{gartner_dev}) and may be replaced to reduce the cost of the maintenance of the old devices, which has a direct impact on authenticator management.
FIDO authenticators usually store private keys in a Restricted Operating Environment, such as TEE based on ARM TrustZone hardware. This environment applies Key Protection Security Measures (SM-1~\cite{FIDOAllliance_sec_ref}), which prevent key export 
and is considered a desirable security feature. However, this security feature also introduces a significant trade-off between security and usability. When a device is replaced, the usability of the platform authenticator (PA) is diminished, 
and employees are required to re-register the authenticator with the system each time a new device is issued. 
Alternatively,  they can use \roaming s such as USB tokens. But even \roaming s can be affected by device replacement, for example, tokens with a USB-A port may be incompatible with USB-C only machines. Additionally, the process of re-registering an authenticator needs to be secured \review{at least as well as} the initial credential binding process. Interestingly, industry experts proposed a solution called multi-device credentials~\cite{FIDOAllliance_multi_device} (a.k.a. passkeys), however, the relaxed security model (e.g., extractable keys) makes them not suitable for the secure enterprise environments. 
\review{

These considerations add complexity to FIDO2 deployment, and as shown in the results of our user study, authenticator migration is perceived as a challenging process.}

\subsection{Authentication}
\noindent \textbf{Telework Authentication:}
We examine how authentication in remote work scenarios is being addressed by FIDO2, following the NIST SP 800-46r2 \cite{SP80046r2} categorization of remote work (telework).

Telework Client Devices such as mobile phones, personal PCs, and laptops usually implement local verification based on credentials such as PIN or biometrics. While mobile devices have already implemented FIDO2, laptops do not yet implement this protocol pervasively (at the time of writing, only the Windows Hello for Business framework provides out-of-the-box FIDO2 support), which makes a FIDO2 rollout challenging in the diverse enterprise environment.

A direct connection with the application server (Direct Application Access), even though simple, is not always an option due to security and regulatory requirements (e.g, applications are inside a tightly controlled perimeter). A common solution is to apply an additional layer of security (and authentication) between the telework client devices and servers. Usually, this is enforced  using tunneling, application portals, or both. Tunneling is often implemented as a VPN (Virtual Private Network), which creates a secure connection between the client and the VPN gateway, thus allowing the client to connect to application servers. 
Alternatively, application portals (e.g., virtual desktop infrastructure) move the application clients into a controlled environment, through which they are accessed. 
In both cases, a client has to authenticate. For both mobile and web applications, FIDO2 provides a well-defined and supported authentication method. This method can be easily implemented for application portals, whereas for VPN clients, the support depends on the vendor's implementation. 

Notably, enterprise authentication is not limited to applications, but to infrastructure as well. Remote access to servers is equally essential to the company's operations, regardless of technology (graphical or command-line-based). In Windows environments, Remote Desktop Protocol (RDP) is commonly used, however, it does not support FIDO2 yet. In the case of Linux servers, Secure Shell Protocol (SSH) is a common form of remote access. At the time of writing, OpenSSH supports FIDO2 only as a way to protect private keys. 

\smallskip
\noindent \textbf{Credential Delegation:}
Typically employees' accounts are bound to a single identity (e.g., through the enrolment process). However, single ownership of an account (and associated credentials) can be insufficient to cover more sophisticated use cases - e.g. multiple individuals may need access to a shared resource and this requires credential delegation. Such delegation is often accomplished by credential sharing which makes it impossible to trace and hold individual users accountable for their actions on that resource. 
There are however other approaches for credential delegation. As described by Grosse et al.~\cite{6381399}, the best delegation approach is one directly implemented in an application (e.g. calendar sharing), however, this approach does not scale. Alternatively, as shown in the AARC project~\cite{AARC-delegation}, delegation can be realized using protocols such as OAuth2.0, yet this approach requires additional infrastructure and is focused on delegating access to a non-human actor and via APIs.

The FIDO2 protocol does not natively support credential delegation - tight binding of user verification (e.g., fingerprint) with the authenticator makes sharing of a FIDO2 credential impractical. However, Frymann et al.~\cite{DBLP:journals/iacr/FrymannGM22} describe the use of the Asynchronous Remote Key Generation (ARKG) primitive to generate credentials in FIDO2 which can be delegated.

\smallskip
\noindent \textbf{Shared Credential:}
Shared credentials are a common use case in enterprises. As outlined by Haber et al.~\cite{Haber2018}, some devices and applications are built with only a single local account, and thus a common practice is to share this credential (e.g., password or private key) among the team members.
FIDO2 was not designed to facilitate the shared credential scenario. However, one could configure the FIDO2 authenticator to be shared by the team (e.g., a single \roaming{} without user verification). While the FIDO2 authenticator provides better security properties than passwords or private keys (e.g., anti-cloning measures), it does not solve the fundamental issues with shared credentials (e.g., lack of accountability).

\smallskip
\noindent \textbf{Privileged Accounts:}
Privileged access is common in enterprise environments. Dedicated Privileged Access Management (PAM) solutions are used to supervise how (privileged) accounts are used \cite{NIST_Privilege}. 
As described by Habel et al. \cite{Haber2020}, PAM systems typically provide the functionality of credential vault, session proxy, and audit register as well as Zero-Trust~\cite{Garbis2021} properties with concepts such as least privilege model or Just-In-Time provisioning. 

FIDO2 in the context of PAM can be considered not only as an authentication to access PAM vaults. For example, dynamically created SSH keys (for privileged accounts) can be secured with a personal FIDO2 token (ecdsa-sk key type), and thus introducing user verification for each use of SSH keys. 
FIDO2 can also facilitate continuous authentication that  could ensure the integrity of privileged sessions routed via PAM proxies. Although continuous authentication is not directly supported in FIDO2, the primitives such as silent authentication (without human action) are already in place. An example of how this could be achieved using FIDO2 extensions was proposed by Klieme et al.~\cite{9343231}. 

\medskip
\review{
Even though the authentication use cases listed above are not exhaustive, they demonstrate the variety of constraints and requirements that an authentication system needs to address. Notably, the popular use cases already pose challenges (see our user study results) and these are usually amplified by custom (i.e., business-specific) and edge case variations. 
}

\subsection{Account Recovery}
Account recovery is the process of regaining access if the primary authentication method cannot be used (e.g., lost or stolen authenticator). Usually implemented as fallback authentication, account recovery has to be done using authentication which is at least as secure as the primary method. Various fallback authentication methods, including their usability and security properties have been evaluated by academics. AlHusain et al.~\cite{ALHUSAIN2021102487} provided an extensive review of fallback authentication research and concluded that the most popular methods are based on mobile devices. An assessment of the usability of fallback authentication was done by Markert et al.~\cite{fallback_authn}. Their preliminary study shows that SMS and email-based methods are more usable than other approaches. 

Workforce account recovery procedures significantly vary from the customer experience. As outlined by Saxe \cite{enterprise_ar}, enterprise processes focus on security and access continuity. Additionally, the reduced user base (i.e., only personnel) and usability requirements, allow leveraging human-based solutions (e.g., help desk or peer checks). Even though such methods are not flawless (e.g., prone to social engineering), their flexibility and non-programmatic nature contribute to access continuity. Interestingly, as shown by Reynolds et al.~\cite{255254}, over 16\% of help desk tickets in their study were related to account recovery.

The security and privacy guarantees of FIDO2 pose a real challenge to implementing secure and usable recovery solutions (also indicated by our respondents, see Section \ref{sec:usabilit_study}). The official FIDO Alliance recommendations \cite{FIDOAlliance_lifecycle} state that recovery can be done in a self-service manner using strong fallback authentication 
or an identity vetting procedure. Kunke et al.~\cite{arxiv.2105.12477} evaluated 12 account recovery methods found in academic publications and in the wild (following the Bonneau et al.~\cite{6234436} framework) and concluded that FIDO backup tokens are the best of available methods. Notably, selecting a weaker recovery method opens the gate for FIDO2 downgrade attacks described by Ulqinaku et al. \cite{274610}. Similarly to the migration use case, FIDO2 multi-device credentials \cite{FIDOAllliance_multi_device} can be used for account recovery. 
An interesting solution was proposed by Visa Research~\cite{cryptoeprint:2022/1555}, where a managerless signature group can be registered instead of a single authenticator, thus improving the usability of the multi-authenticator approach to account recovery.

\subsection{Deprovisioning}

The deprovisioning process marks the end of the identity lifecycle. Usually, user's access is revoked and the account is deleted or suspended. Simultaneously, the account's active sessions are terminated and all bound authentication factors are deregistered. In the context of FIDO2, deregistration is as simple as removing the public key from the server storage. For the server-side credentials, storage clean-up is enough, however, in the case of discoverable credentials, they remain in the authenticator memory. Even though invalid, they take up space which, is an issue for devices with limited storage (e.g., 25 keys for some Yubico hardware tokens). Before the release of CTAP 2.1 the only option to remove a key was to reset the authenticator, which removed all created keys. With the latest protocol version, an \textit{authenticatorCredentialManagement} API was added to enable the removal of a single key. This API is an authenticator side operation that has to be triggered by the user via a CTAP client (e.g., specialized software provided by the device vendor). 

\review{
In contrast to other procedures such as enrollment or authentication, deprovisioning may be considered rudimentary. Nonetheless, it was still included in the list of challenging processes according to our user study.
}

\subsection{Usability}
Regarding workforce usability, the integration decisions impact how FIDO2 is perceived. For example, the majority of modern devices already have built-in \platform s, and thus employees' private devices can be leveraged ( e.g., when the ``bring your own device'' policy is allowed). This approach was researched in the academic institution context by Weidman et al.~\cite{10.1145/3134600.3134629} who found that employees perceived using private devices as unprofessional. 
On the other hand, workforce FIDO2 deployment in a small company was studied by Farke et al.~\cite{255646}.
Their findings indicate that most of the employees found key-based login as usable, however, several of them were unconvinced of security benefits and found password managers integrated with web browsers faster to use. Similarly, Farke et al.~\cite{281252} examined passwordless authentication using Windows Hello for Business usability in the company setup and found it faster, more responsive, and convenient to use. 

As mentioned before end-user usability is not the only one to consider. 
Ease of deployment and post-deployment maintenance plays a significant role in the solution being successful. In particular, security-related products come with a demanding configuration process (e.g., Krombholz et al.~\cite{203690} revealed TLS configuration issues, some of which are challenging even for security experts). This applies to FIDO2 as well, and like any technology it struggles with the challenges of the early adoption stage (e.g., lack of technical know-how and examples). As noted by Alam et al.~\cite{10.1145/3319535.3363283}, 
the ongoing development of WebAuthn features and tools is critical. In recent years, the range of available tools and libraries noticeably increased (see WebAuthn resources).\footnote{https://github.com/herrjemand/awesome-webauthn} Academics have also contributed to the development aspects of FIDO2. For example, Grammatopoulos et al.~\cite{10.1145/3465481.3469209} provided an analysis tool for FIDO2 traffic. Regarding deployment, FIDO2 authentication is currently available in all major cloud providers, however, as noted by Gordin et al.~\cite{9638271}, not all solutions provide an easy FIDO2 integration (e.g., Open Stack).

\review{Despite the admirable work of the FIDO2 community to educate about the protocol, usability as well as technical knowledge is still a major problem for FIDO2 implementers (see Figure \ref{fig:fido_questions}). }

\section{FIDO2 in Enterprise Usability Study}\label{sec:usabilit_study}
To answer the study question, we conducted a user study, collecting feedback from IT professionals with practical experience in IAM. 
\rev{We collected complete responses from 118 participants, all of which were included in our analysis. Only one free-text response was excluded as it was not interpretable (i.e., single-letter response). }

\subsection{Study Design and Methodology}
We prepared an online questionnaire following the human-computer interaction research guidelines described by Lazar et al. \cite{LAZAR2017105}. The online format provides us the flexibility to target our interest group remotely and globally. The questionnaire was written in English and implemented as a publicly available and anonymous Google Form. The questions for the study were carefully designed by an academic researcher and senior IAM subject matter expert based on knowledge, experience, and literature review. Then, we run an internal evaluation of questions through the review process carried out by three independent senior academic researchers, and a pilot test among our internal research group (15 subjects).


The questionnaire consists of 26 open and multiple-choice questions grouped into four sections. To allow unaided answers, we provided an "Other" option where appropriate. 
The full questionnaire can be found on our repository \footnote{https://bit.ly/fido-survey}. Depending on the question's nature, we used the following statistical tests to determine relations between variables in our study: Cramér’s V ($\chi^2$), Kruskal-Wallis ($KW$) and Pearson correlation. Unless otherwise indicated, Cramér’s V was used. Presented results have small effect unless otherwise noted. Results with $p\leq0.05$ were considered statistically significant
\ifx\version\sel
\footnote{\rev{See Appendices~\ref{sec:kruskal},~\ref{apx:survey_relations} for adjustments for multiple hypothesis testing.}}. 
\else
\footnote{\rev{See full version of the paper for adjustments for multiple hypothesis testing.}}. 
\fi
\rev{The answers to the open-ended questions were grouped and summarised by one of the authors. }
We describe our findings in the sections below and provide an exhaustive list in 
\ifx\version\sel
Appendices~\ref{sec:kruskal} and \ref{apx:survey_relations}.
\else
the full version of the paper.
\fi

\subsection{Recruitment and Participants}
We opened the user study in the first quarter of 2022 and performed an advertising campaign throughout the year. 
\rev{Participation in the study was voluntary and did not include compensation.}
The study target group is individuals working towards implementing cybersecurity solutions. Importantly, to increase the completeness of the study, we invited professionals from different roles including decision-makers \rev{(e.g., C-level executives, directors, department heads)}, managers, subject matter experts (SMEs), and developers. The call for participation was promoted through leaflets, personal appearances at cybersecurity conferences and meetups, personal contacts, public posts on social media such as LinkedIn, related discord channels, and the fido-dev email list. \rev{For our analysis, we considered only complete submissions of the study questionnaire.}

We acknowledge that our recruitment approach introduces a bias towards professions and organizations particularly interested in FIDO2. However, as the goal of the study is to learn the FIDO2 perception from practitioners (not the perception of FIDO2 in the wild), we claim that this bias does not adversely impact our results. 

\subsection{Ethical Considerations}
The study was approved by the university's ethical review board. We endeavored to ensure that collected data does not contain any private or sensitive information. Participation in the study was voluntary and we did not require or record any form of identification. All participants were informed about the study terms before engaging with the questionnaire.

\begin{figure}[t!]
    \centering
    \includegraphics[width=\linewidth]{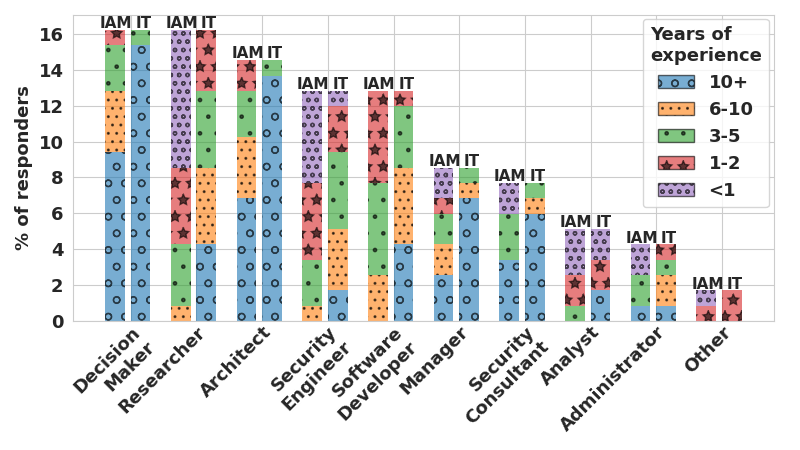}
    \caption{\small Years of experience in Identity and Access Management (IAM) and Information Technology (IT) by profession.}
    \label{fig:full_exp}
\end{figure}


\begin{figure*}[t!]
    \centering
    \includegraphics[width=\linewidth]{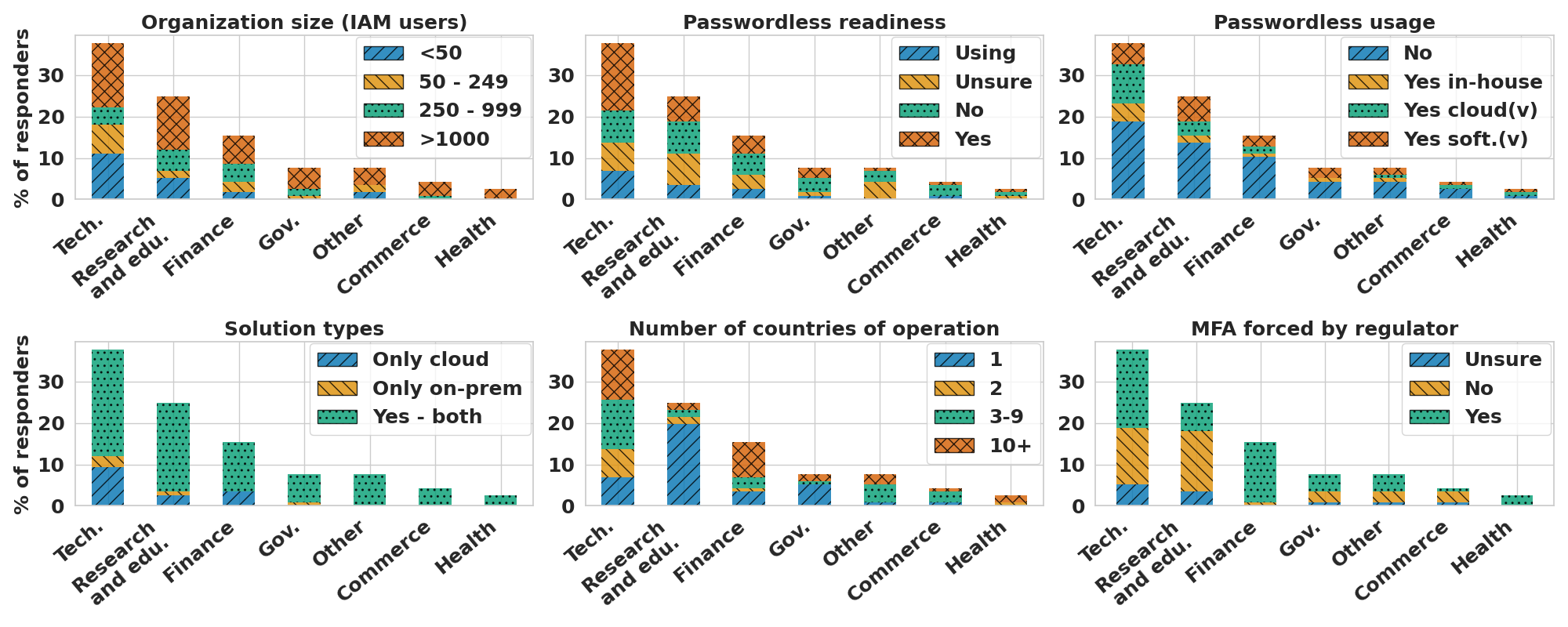}
    \caption{\small{Organization and passworldess authentication related responses by industry.}}
    \label{fig:org1}
\end{figure*}

\subsection{Participants Profile}
\rev{Our study goal is to understand the professional opinions of our respondents. Therefore, we have opted not to include demographic questions in the profile section. We justify this decision due to concerns about the privacy of our respondents and their hesitancy to share personal information. Instead, our focus has been on gathering data that delineates the professional profiles of our respondents.}
As shown in Figure \ref{fig:full_exp}, our participants hold a diversity of roles in the enterprise hierarchy (no role had more than 16.2\% of responses). Over 50\% of the respondents identified themselves as senior IT personnel (i.e., 10+ IT experience) and being involved in IAM for at least 3 years. Moreover, over 40\% of the respondents actively participate in an IAM project at their organization. The roles diversity, strong IT and IAM background and hands-on experience of the participants support the credibility of our data. In regards to professions, the frontliners in investigating and adapting new technology (i.e., decision makers, researchers, and architects) were the most numerous groups. Moreover, we observed that a significant proportion of respondents are technical and strategic leaders (decision makers, architects, and managers). They stand for over 40\% of answers and qualify as highly experienced specialists with over 90\% having more than 10 years experience in IT and 50\% more than 10 years in IAM.

The study participants represent a diverse spectrum of organizations including large global enterprises. Over 50\% of respondents work in organizations that provide IAM services for more than 1,000 users and over 35\% of them are multi-region companies (i.e., operating in 2 or more countries). In total, over 60\% of the data set contains multi-country organizations. In terms of industries, over 75\% of responses were submitted by employees of three sectors: Technology, Research and Education, and Finance and over 50\% answers indicate companies operating in regulated environments (e.g., PCI DSS or HIPAA). Interestingly, the majority of organizations (almost 80\%) use both cloud and on-premise applications. The above observations and data presented in Figure~\ref{fig:org1} strongly suggest that a significant proportion of our participants operate in highly demanding environments which impose additional requirements for authentication.

\subsection{Passwordless Authentication}

\begin{figure}[h!]
    \centering
    \includegraphics[ width=\linewidth]{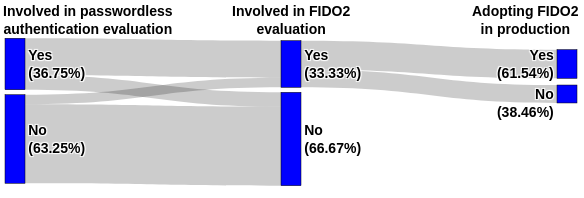}
    \caption{\small Proportions of respondents evaluating passwordless authentication and FIDO2, and adopting FIDO2 in production.}
    \label{fig:evaluation}
\end{figure}

According to our respondents, only 45\% of their organizations are already using or are ready for migration to passwordless authentication. The proportion was highest for the Technology sector (61\%), followed by Finance (44\%) and Research and Education (38\%). Surprisingly, over 30\% declare that their organization is not ready to move away from passwords. In terms of deployments, almost 54\% of respondents admit that their organization does not use any passwordless authentication solution. Among existing production systems the preference splits almost equally between vendor-provided software(18.5\%) and vendor-managed cloud solutions (19.3\%). In-house built solutions were present in only 8.4\% answers. The details for each industry can be found in Figure~\ref{fig:org1}. Almost 37\% of respondents admitted that they were involved in passwordless solution evaluation. One-third participated in the evaluation of a FIDO2 solution. Interestingly, 12\% of respondents who did not evaluate passwordless solutions, answered that they evaluated FIDO2. We hypothesize that FIDO2 may have been evaluated as a 2FA (second factor) in solutions that still rely on passwords as the first factor. Figure~\ref{fig:evaluation} shows an overview of responses on evaluation.

\begin{figure*}[ht!]
    \centering
    \includegraphics[width=\linewidth]{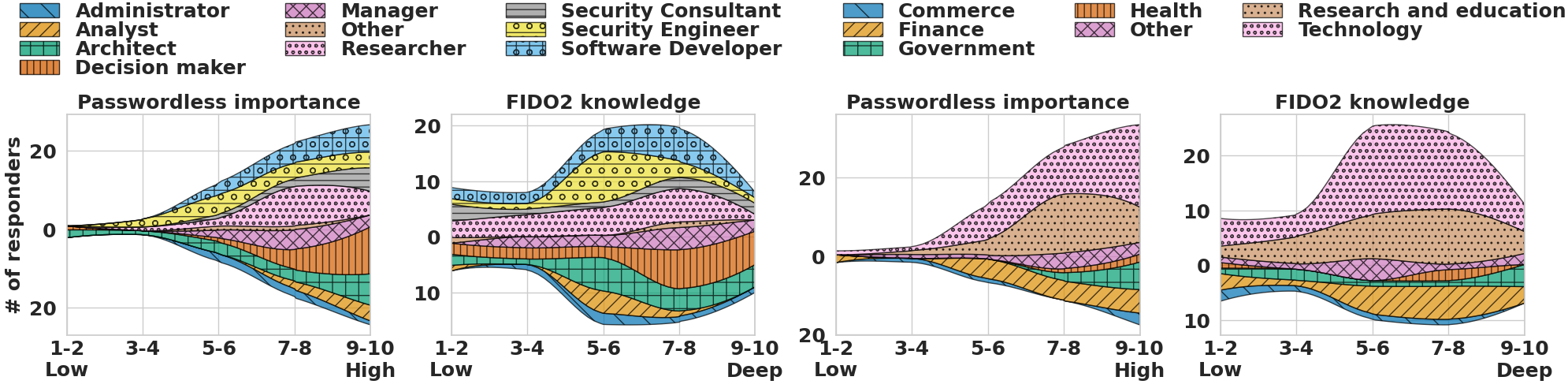}
    \caption{\small{Passwordless authentication importance and FIDO2 knowledge questions by profession.}}
    \label{fig:importance_knowledge}
\end{figure*}

The last question in this section measured the participants' perception of the importance of passwordless authentication on a 10-point Likert scale (see Figure~\ref{fig:importance_knowledge}). Generally, passwordless authentication is perceived as important with the majority of answers above the neutral threshold (i.e., 5 on the scale) and 43.7\% of answers stating it is extremely important. In terms of professions, we noticed a significant increase in answers towards high importance from decision makers, security consultants, and architects. We hypothesize that these roles need to follow new technologies and threats to direct work in their organizations, and thus are more convinced and aware of the importance of passwordless authentication. Interestingly, managers' and researchers' answers were less extreme. We speculate that their role involves consideration of a wider context beyond IAM, 
which might decrease the apparent importance of passwordless authentication. Surprisingly, the analysis by industry (right two panels of Figure~\ref{fig:importance_knowledge}), showed that government employees are highly convinced of the importance of passwordless authentication. 

Regarding statistical analysis, we identified the following relations. Managers were found to not confirm the passwordless readiness of their organizations, whereas decision makers more frequently answered positively. Similarly, the technology industry stands out in indications of passwordless readiness. Organizations operating in non-regulated environments were linked with not being ready for passwordless. Passwordless importance was found to be related to readiness ($KW$ small effect). Experienced employees (IT 10+) and in particular decision makers, architects, and security consultants were found to more frequently use passwordless solutions. Similarly, the experienced personnel more frequently participated in the evaluation of passwordless solutions. Surprisingly, participants with less than one year of experience as well as those working for the government tended to have been involved in passwordless evaluation.







\subsection{FIDO2}\label{sec:fido_study}
The FIDO2 part of the study starts with knowledge evaluation. The distribution (presented in Figure~\ref{fig:importance_knowledge}) peaks exactly in the middle with a median of 6. We analyzed the knowledge distribution per profession and industry and found that the medians of management roles (i.e., decision maker and manager) are 1 point above the distribution median. Interestingly, we observed lower medians (5) for the engineering roles (i.e., software developer, security consultant, and analyst). In terms of industries, the employees of typically regulated industries recorded higher median values: Government (7), Health (7), and Finance (6.5). The lowest median of 4 was found for the Commerce industry. Passwordless importance and FIDO knowledge show a positive correlation ($\rho = 0.2, p=0.03$). The FIDO knowledge variable was found to relate to IAM experience with a large effect ($KW$), and to having evaluated passwordless (medium effect) and FIDO (large effect) ($KW$). Interestingly, a medium effect relation was also found with the number of countries where an organization operates ($KW$), and a large effect relation with ongoing FIDO deployments in production ($KW$).


Respondents who participated in the evaluation of any FIDO2 solution were asked additional, more detailed questions. A total of 39 participants 
(33\%) qualified for this section. Almost half of the respondents are decision makers and architects, and almost 80\% are highly experienced in IT (10+ years) and over 40\% in IAM. Interestingly, 62\% of this group is in the process of adopting FIDO2 in production (see Figure~\ref{fig:evaluation}). The data analysis shows medium effect relations between participants experienced in IAM and FIDO2 evaluation, however, the inverse was found for employees in IT with 10 or more years of experience. Evaluation of FIDO was found to relate to the respondents from the government sector. 
Notably, we found relations with ongoing FIDO2 deployment, many of which have a medium or large effect. Firstly, the relation with top experienced IAM personnel had a large effect, whereas participants with 1-2 and 3-5 years of experience had a medium effect relation to \emph{not} having FIDO deployment in production. Interestingly, decision makers were the only profession found to have a relation with FIDO deployment in production (medium effect) and only one use case, "privileged accounts", showed a relation. A relation to not having a production deployment was found with single-country organizations and the research and education industry (both with medium effect).



\begin{figure*}[ht!]
    \centering
    \includegraphics[width=\linewidth]{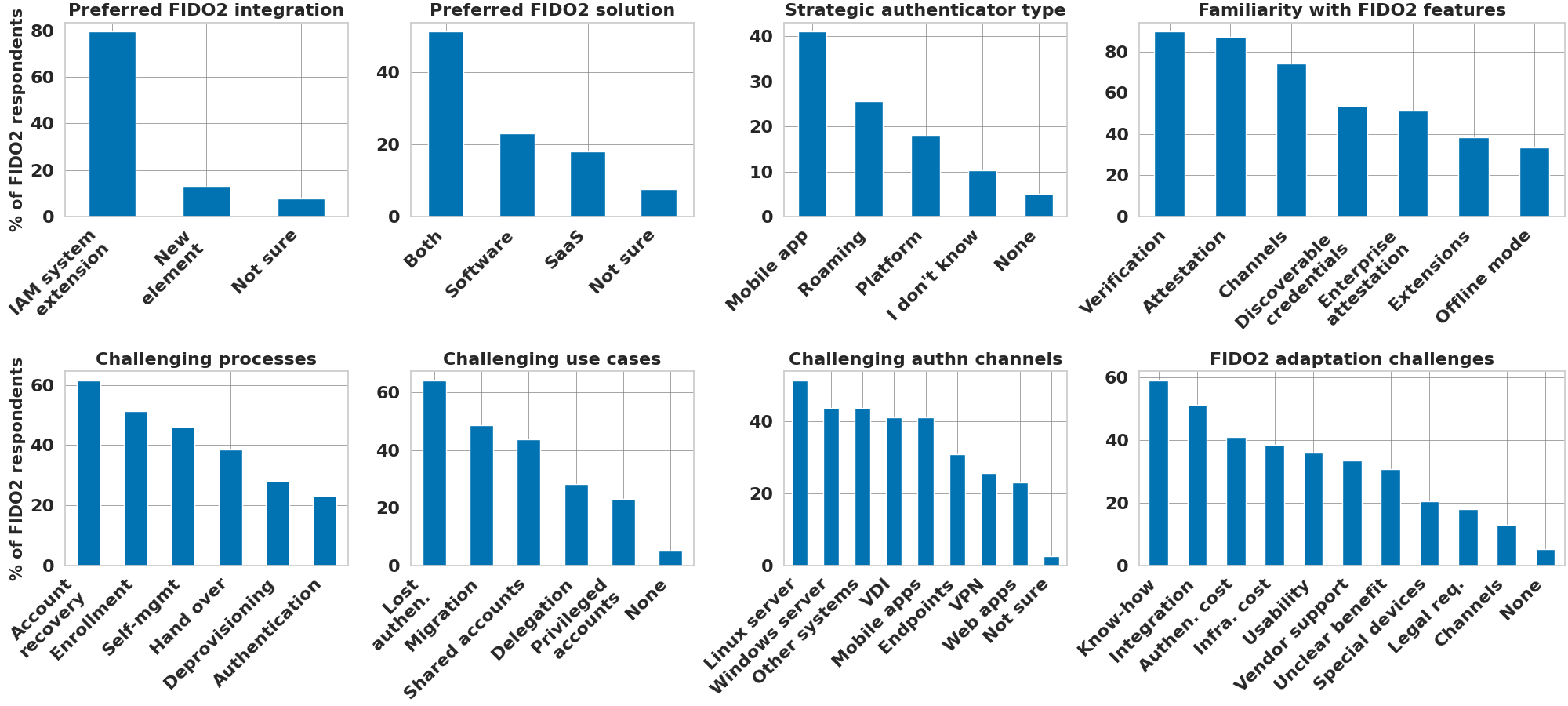}
    \caption{\small{FIDO2 preferences and challenges (only respondents involved in FIDO2).}}
    \label{fig:fido_questions}
\end{figure*}

Figure~\ref{fig:fido_questions} summarises responses to questions on FIDO2 preferences, familiarity, and challenges. 
The respondents show a clear preference for FIDO2 integration as an existing IAM solution extension (79\%). In terms of the deployment model, 51\% of participants prefer to have both software and vendor-managed solutions (e.g., in cloud), with 23\% preferring software only and 18\% purely SaaS (Software as a Service). Interestingly, the SaaS option 
had a medium effect relation with the technology sector, participants with less than 1 year of experience, organizations with on-prem and cloud, and those with only cloud presence. Regarding the preferred authenticator, we observed mobile applications as the most desired type of authenticator (41\%). The second pick was roaming authenticators (26\%) followed by platform authenticators (18\%). Interestingly, hardware roaming authenticators had a medium effect relation with staff with 1-2 years of IT experience and with organizations below 50 users. On the other hand, platform authenticators were related (medium effect) to organizations not yet ready to migrate and those with on-prem presence only. Interestingly, we found a few medium-effect relations to ``I don't know which authenticator type'' and ``Neither authenticator type'' on authenticator preferences. For example, the technology industry, the architect profession, and the deployment of both on-prem and cloud. This might suggest insufficient knowledge about authenticator types. 

In addition to preferences, we measured our participants' familiarity with FIDO2 features. Clearly, user verification and attestation are well-known features with 90\% and 87\% of responses respectively. A slightly less recognized feature of FIDO (74\%) is the support for various transport channels. Surprisingly, two features particularly useful in the enterprise context: discoverable credentials (a.k.a. resident keys) and enterprise attestation, are only known to about half of the respondents (54\% and 51\%). The least known features of FIDO2, according to our respondents, are FIDO extensions (38\%) and Offline authentication mode (33\%).

To understand FIDO2 challenges, we asked our participants which aspects of passwordless FIDO2 authentication are considered difficult. We organized questions into four categories: adaptation, processes, use cases, and authentication channel challenges. 
For the questions in this group, participants could choose to submit open responses instead of the suggested responses, which we discuss in Appendix \ref{sec:openanswers}. 

The main reason why FIDO2 adaptation is challenging according to our respondents is a lack of knowledge and know-how (59\%). Over half of the participants (51\%) identified integration with their systems as being complicated. Interestingly, for the third and fourth place, the respondents selected the cost of authenticators and infrastructure (41\% and 38\% respectively). FIDO2 usability is considered a challenging factor for over one-third (36\%) of participants and 33\% believe that vendor's support is insufficient. Surprisingly, we found that the unclear benefits of FIDO2 authentication are a noticeable adaptation challenge (31\%). The requirement for a specialized FIDO2 device (i.e., an authenticator) is a challenging factor for only 20\%. Legal requirements and connectivity channels were identified as the least challenging (18\% and 13\% respectively). Statistical analysis shows that the "no challenges" answer relates to less experienced participants (below 1 year of IAM experience with a large effect and in 1-2 years in IT). Similarly, a relation was found between the security consultant role and SaaS as the preferred FIDO2 deployment model. Interestingly, connectivity channels as a challenge were found related to single-country organizations and participants who are unsure about the preferred deployment model. Furthermore, we found a relationship between the challenge of legal requirements with participants with 1-2 years of experience in IT.


For the identity lifecycle processes, FIDO2 was found the most challenging for account recovery (62\%). Over half (51\%) of respondents marked the enrollment flow, and 46\%  found self-management with FIDO2 challenging. The initial and final processes (i.e., hand-over and deprovisioning) were only selected by 38\% and 28\% participants respectively. Surprisingly, the main functionality of FIDO2 (i.e., authentication) was identified as problematic to integrate into the processes for only 23\% respondents. The only significant relationship found for challenges with identity lifecycle was between authentication processes and organizations below 50 users. 

Regarding challenging use cases, the majority of respondents (64\%) selected ``lost authenticator'', followed by ``authenticator migration'' (49\% of answers). Similarly, shared accounts were identified as a challenge by 44\%. Interestingly, less than 30\% of respondents believe that authentication delegation (28\%) and privileged account (23\%) use cases are challenging. 
Statistical analysis found that the ``none'' answer was related to software developers, organizations operating in multiple countries (3-9) (large effect) and participants with 3-5 years of IAM experience.  The delegation use case showed a relation with the research and education industry. Interestingly, the migration use case had a relation to uncertainty about passwordless readiness, and challenges around privileged accounts were found to be related to experienced IAM employees and participants working on the production deployments of FIDO2.


The last questions' category focuses on the challenging authentication channels. Interestingly, authentication to remote servers is perceived as problematic. Linux servers were identified as the most challenging environment (51\%), followed by Windows (44\%) and other servers (44\%), with only a few percentage points less (41\%) for both VDI and mobile applications. Access to endpoints was identified as challenging by 31\% of participants and VPN authentication by 25\%. Surprisingly, the main target of FIDO2 authentication (i.e., web applications) was selected by 23\% of the respondents. Statistical analysis found medium effect relations between the "not sure" answer and both managers and participants with 6-10 years of IAM experience. Additionally, a relation was found with software as a preferred deployment model. Interestingly the VPN channel was related  to participants with 3-5 years of IT experience and participants unsure about their company's readiness for a passwordless migration. Furthermore, the mobile channel was related to organizations that already are using a passwordless solution.

\subsection{Free Text FIDO2 Questions}\label{sec:openanswers}
Considering the wide scope of possible challenges with FIDO2 adoption, 
we provided an option to submit free text answers which we discuss below. 
Regarding the challenges to FIDO2 adoption, respondents most frequently addressed the incompleteness of the FIDO2 environment. Firstly, technical issues were drawn to our attention. The following opinions: \textit{``missing/incomplete FIDO2 mobile solutions''}, \textit{``sync between multiple platforms (ms/apple/google)''}, \textit{``lack of authenticator support and availability of authenticators with biometrics''} suggest that the existing implementations do not meet industry expectations, and thus make it challenging to roll out a FIDO2-based solution. Similarly, one of the respondents addressed technical challenges with the edge cases (\textit{``support legacy applications and use cases where USB/NFC/BLE cannot be used''}). Interestingly, we received an observation about FIDO2 documentation (\textit{``limited documentation for fido2 server implementors''}) and challenges in the correct delivery method (\textit{``secure delivery of a roaming authenticator''}).

Respondents spotted challenges in aligning organizations and users towards FIDO2. In particular, two answers (\textit{``competing priorities''} and \textit{``unclear benefit of FIDO2 vs Push Notifications''}) suggest that the idea and importance of secure phishing-resistant passwordless authentication are not yet commonly known. Interestingly, complex enterprise environments can impact how challenging the FIDO2 integration is. For example, one of the respondents noted that their organization has a \textit{``requirement for both mobile devices and dedicated authenticators''}. Regarding end users, user experience was indicated as a challenging factor. One respondent identified user interface messaging as an issue (\textit{``it is confusing for end users at first, all these different prompts because different browsers present differently.''}). The second opinion described a specific flow that lacks smooth user experience (\textit{``when integrating with OpenID Connect, the user experience involves a switch from mobile app to browser for authentication''}).







The responses collected for organizations' most important FIDO2 use cases revealed that completeness and unification of the FIDO2 environment play a critical role. In particular, the support for various authentication channels was repeatedly mentioned (\textit{``VPN, access to a web app, mobile app, access to servers, critical infrastructure like DNS, G Suite, AWS''}, \textit{``access to desktop computers and laptop devices; Access to VDI''}). Additionally, respondents outlined a unified user experience as a significant factor in their organizations (\textit{``same/similar UX ... as password managers, which users are familiar with''}, \textit{``FIDO2 needs to be ubiquitous across platforms...''}). Another use case identified by our participants is device support and BYOD (bring your own device) policy. For example, \textit{``ease of use through manufacturer level support on all employee devices, ...''} as well as \textit{``FIDO2 BYOD solutions for mobile''} are considered to be crucial use cases. Surprisingly, only one opinion mentioned security features as important use cases (\textit{``phishing resistant MFA''}).







In the final open question, we captured participants' opinions about other (i.e., not mentioned before) obstacles and challenges that could prevent or negatively impact FIDO2 deployment. Interestingly, the majority of our respondents' answers point to human-related issues. The first one addresses user adaptation challenges. For example, \textit{``the resistance of the users/administrators to new technology''}, \textit{``user awareness''} and \textit{``relatively new, ... it’s hard for most people to get over the hump benefit of PKI with binding on the authenticator''} suggest that understanding of FIDO2 technology is not sufficient, and thus difficult to successfully incorporate in the organization's security suite. Moreover, the following responses \textit{``lack of a champion in a leadership role''} and \textit{``complicated for non-technical users to understand''} reveal that organizations struggle with passwordless migration process due to missing know-how. Two of the respondents expressed security concerns related to the quality of authenticators (\textit{``bugs due to local software errors on devices, producing inconsistencies in authentication reliability''}) and the latest advancement in the FIDO2 environment (\textit{``Apple's implementation of Passkeys is a concern without DPK or attestation...''}). Interestingly, one of the respondents argued that some organizations already use passwordless solutions and do not see incentives in migrating to FIDO2 (\textit{``US Federal Government already has the PIV/CAC. They don't see much value in FIDO2, yet''}).

\review{\section{Study Findings}


Statistical analysis as well as free text answers provide a clear picture of the challenges with FIDO2 integration. One of the major factors that discourage FIDO2 adaptation is know-how and toolset. Our respondents believe that even though their general knowledge of FIDO2 is satisfactory, the practical knowledge and integration paths are still missing and pose a major obstacle. In particular, an adequate handling of processes such as account recovery or enrollment, and edge cases like lost authenticator were found challenging. Interestingly, fundamental processes such as registration and authentication were also found to pose significant challeges by the study. Solution costs and missing native integrations were spotted as a major adaptation blockers. In particular, integration with commonly used servers (e.g., Windows or Linux) poses a significant challenge for enterprise deployment. Furthermore, the respondents clearly articulated that usability, more precisely, differences in the presentation (UI) and process (UX) pose a challenge for the integration into employees' daily routines. 

Apart from challenges, our study identified the most preferred FIDO2 adaptation model (i.e., an extension of an existing IAM platform, which can run as both SaaS and a software solution) as well as the preferred FIDO2 authenticator (i.e., mobile application). We learned that even though respondents are familiar with the fundamental FIDO2 features, the more advanced properties are less known, which could explain why the unclear benefits were perceived as a challenge.

\section{Conclusion}
In this work, we asked IT specialists and FIDO2 implementers about their perspectives on the challenging aspects of FIDO2 adaptation. Our initial review of use cases shows that the FIDO2 ecosystem, even though having an admirable coverage of the use cases, still lacks solid and production-ready solutions for certain processes (e.g., account recovery). Our findings were confirmed by the user study, in which we managed to identify and order the most challenging aspects of FIDO2 adaptation (i.e., missing know-how, cost, and usability). We believe that our results, backed by the specialists' feedback, provide clear directions for future FIDO developments. In particular, we hope our study will be used as a guide for usability researchers as well as the FIDO community including standardisation bodies (e.g., FIDO Alliance), commercial vendors, and enterprises. 
}





\bibliographystyle{plain}
\bibliography{./usenix2022_SOUPS.bib}

\appendices

\ifx\version\sel

\begin{table}[h!]
\section{Kruskal-Wallis relations} \label{sec:kruskal}
    \centering
    \begin{tabular}{l|c|c|c|c|c|c}

{} &  Q1 &  Q2 &     p &      V &  effect & LSU$^*$ \\
\hline
5 & 15 & 14 & 0.000 & 46.035 & large & T \\
4 & 13 & 14 & 0.000 & 26.086 & large & T \\
1 & 3 & 14 & 0.000 & 21.965 & large & T \\
3 & 7 & 14 & 0.006 & 12.348 & medium & F \\
2 & 4 & 14 & 0.012 & 6.331 & small & F \\
6 & 18 & 14 & 0.022 & 5.227 & medium & F \\
0 & 10 & 12 & 0.035 & 8.583 & small & F \\
\hline
\end{tabular}

    \caption{\small{Kruskal-Wallis test results ($p<0.05$). See Appendix~\ref{apx:survey-questions} for question details.\\\hspace{\textwidth} $^*$The (non-adaptive) one-stage linear step-up procedure (LSU) for controlling the false discovery rate.}}
    \label{tab:my_label}
\end{table}

\fi

\label{sec:experience_full}
\begin{figure}[h!]
    \centering
    \section{All Participants experience} 
    \includegraphics[width=0.8\linewidth]{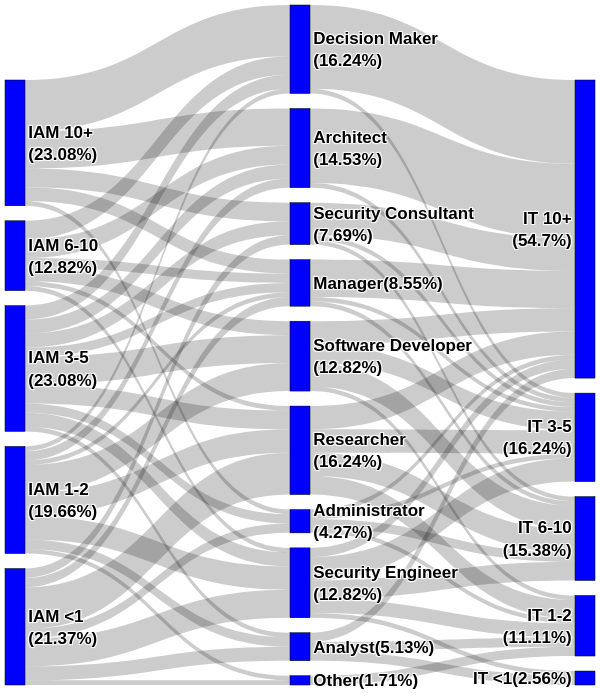}
    \caption{\small{Distribution of experience in Identity and Access Management (IAM) and Information Technology (IT) of respondents' professions: 
    }}
    \label{fig:fido_participants}
\end{figure}


\label{sec:fido_participants}
\begin{figure}[h!]
    \section{FIDO2 Participants} 
    \centering
    \includegraphics[width=0.7\linewidth]{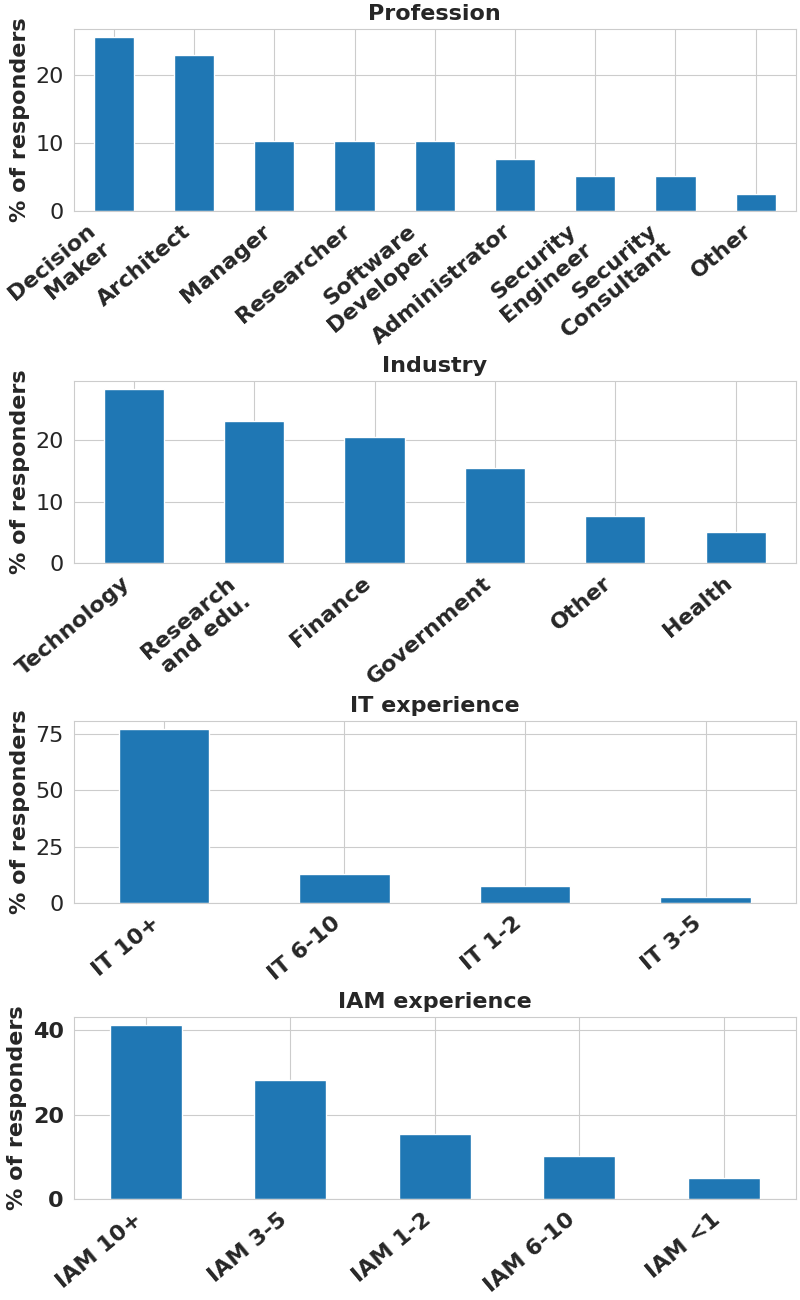}
    \caption{\small{Profile of respondents answering FIDO2 section.}}
    \label{fig:fido_participants_profile}
\end{figure}

\clearpage

\ifx\version\sel

\onecolumn


\section{User Study Participants}
\begin{table}[h!]
    \centering
\begin{tabular}{l|c|c|c|c|c|c|c|c|c|c|c}
& \rot{90}{\vtop{\hbox{\strut Decision}\hbox{\strut Maker}}} & \rot{90}{Researcher} & \rot{90}{Architect} & \rot{90}{\vtop{\hbox{\strut Security}\hbox{\strut Engineer}}} & \rot{90}{\vtop{\hbox{\strut Software}\hbox{\strut Developer}}} & \rot{90}{Manager} & \rot{90}{\vtop{\hbox{\strut Security}\hbox{\strut Consultant}}} & \rot{90}{Analyst} & \rot{90}{Administrator} & \rot{90}{Other} &    \rot{90}{Sum} \\ \hline
Total    &                    16.24 &                16.24 &               14.53 &                       12.82 &                        12.82 &              8.55 &                          7.69 &              5.13 &                    4.27 &            1.71 & 100.00 \\ \hline
IT 10+   &                    15.38 &                 4.27 &               13.68 &                        1.71 &                         4.27 &              6.84 &                          5.98 &              1.71 &                    0.85 &            0.00 &  54.70 \\ \hline
IAM 10+  &                     9.40 &                 0.00 &                6.84 &                        0.00 &                         0.00 &              2.56 &                          3.42 &              0.00 &                    0.85 &            0.00 &  23.08 \\ \hline
IT 6-10  &                     0.00 &                 4.27 &                0.00 &                        3.42 &                         4.27 &              0.85 &                          0.85 &              0.00 &                    1.71 &            0.00 &  15.38 \\ \hline
IAM 6-10 &                     3.42 &                 0.85 &                3.42 &                        0.85 &                         2.56 &              1.71 &                          0.00 &              0.00 &                    0.00 &            0.00 &  12.82 \\ \hline
IT 3-5   &                     0.85 &                 4.27 &                0.85 &                        4.27 &                         3.42 &              0.85 &                          0.85 &              0.00 &                    0.85 &            0.00 &  16.24 \\ \hline
IAM 3-5  &                     2.56 &                 3.42 &                2.56 &                        2.56 &                         5.13 &              1.71 &                          2.56 &              0.85 &                    1.71 &            0.00 &  23.08 \\ \hline
IT 1-2   &                     0.00 &                 3.42 &                0.00 &                        2.56 &                         0.85 &              0.00 &                          0.00 &              1.71 &                    0.85 &            1.71 &  11.11 \\ \hline
IAM 1-2  &                     0.85 &                 4.27 &                1.71 &                        4.27 &                         5.13 &              0.85 &                          0.00 &              1.71 &                    0.00 &            0.85 &  19.66 \\ \hline
IT $<$1    &                     0.00 &                 0.00 &                0.00 &                        0.85 &                         0.00 &              0.00 &                          0.00 &              1.71 &                    0.00 &            0.00 &   2.56 \\ \hline
IAM $<$1   &                     0.00 &                 7.69 &                0.00 &                        5.13 &                         0.00 &              1.71 &                          1.71 &              2.56 &                    1.71 &            0.85 &  21.37 \\
\end{tabular}
    \caption{\small{Detailed experience of survey participants (percentages).}}
    \label{tab:full_participants}
    \vspace{-2em}
\end{table}


\section{Usability Study Questions}
\label{apx:survey-questions}
\begin{table}[h!]
\centering
    \begin{tabular}{l}
\hline
                                                                                                                                                                                                                                         Questions \\
\hline
                                                                                            1. What is your role within your organisation? \\
                                                                                               2. What is your experience in the IT field? \\
                                                             3. What is your experience in the Identity and Access Management (IAM) field? \\
                                                                 4. Are you currently involved in an IAM project within your organisation? \\
                                                                                             5. What is the industry of your organisation? \\
                                   6. How big is your organisation in terms of number of users (employees, contractors, etc) that sign in? \\
                           7. Is your organisation (along with its customers and users) present in multiple countries? How many countries? \\
                          8. Does your organisation operate in a regulated environment (...) that requires the use of MFA or passwordless? \\
                                                                          9. Does your organisation use on-premise and cloud applications? \\
                                    10. In your opinion, is your organisation ready to move from passwords to passwordless authentication? \\
                                                  11. Is your organisation already using an existing passwordless authentication solution? \\
                                                        12. In your opinion, is it important to adopt a fully passwordless authentication? \\
                                           13. Have you been involved in evaluation (...) of any solution for passwordless authentication? \\
                                                               14. How well do you know FIDO2 technology for  passwordless authentication? \\
                                            15. Have you been involved in evaluation (...)  of any solution based on the FIDO2 technology? \\
                                16. What is your preferred way of implementing FIDO2 passwordless authentication within your organisation? \\
                                          17. What is your preferred model to introduce FIDO2 passwordless solution: software or ... SaaS? \\
                                              18. Are you currently adopting a solution based on the FIDO2 technology ... (in production)? \\
                                              19. Which of the following FIDO2 passwordless authentication features are you familiar with? \\
               20. What are the key ... challenges which you observed during ... adoption of a FIDO2 passwordless authentication solution? \\
  21. Which part of the user ...related process is difficult to change ... to work with FIDO2 passwordless authentication ...? \\
                                 22. Which scenarios do you consider most challenging for adoption of a FIDO2 passwordless authentication? \\
                                                               23. Which authenticator type is considered strategic for your organisation? \\
                              24. In your opinion, which authentication channels can be problematic for FIDO2 passwordless authentication? \\
              25. Please briefly describe the use cases that are most important ... to adopt FIDO2 passwordless authentication?  \\
26. ... what other obstacles (...) have you observed that would prevent ... deployment of FIDO2 passwordless authentication ...? \\
\hline
\end{tabular}
\caption{Usability study questions.}
\end{table}

\vfill\eject
\section{Cramér’s V Relations}
\label{apx:survey_relations}

\begin{table}[h!]
\centering
    \begin{tabular}{l|l|l|l|l|c|c|c|c|c|c|c|c}

\multirow{2}{*}{} &  \multirow{2}{*}{Q1} &  \multirow{2}{*}{Q2} &               \multirow{2}{*}{Q1 val} &            \multirow{2}{*}{Q2 val} &     \multirow{2}{*}{p} &     \multirow{2}{*}{V} &  \multirow{2}{*}{effect} &  \multirow{2}{*}{LSU$^*$} & \multicolumn{4}{|c}{Confusion matrix}  \\
  & & & & & & & & & TP &    FP &    FN &    TN \\
\hline
0  & 3  & 20 & Less than one year  & No challenges    & 0.000 & 0.625 & large  & T & 0.400 & 0.600 & 0.000 & 1.000 \\
1  & 8  & 22 & I'm not sure        & None             & 0.000 & 0.703 & large  & T & 1.000 & 0.000 & 0.012 & 0.988 \\
2  & 1  & 22 & Software Developer  & None             & 0.000 & 0.621 & large  & T & 0.400 & 0.600 & 0.000 & 1.000 \\
3  & 7  & 22 & Yes (3-9 countries) & None             & 0.000 & 0.518 & large  & T & 0.286 & 0.714 & 0.000 & 1.000 \\
4  & 6  & 23 & 250 - 999           & None             & 0.000 & 0.606 & large  & T & 0.400 & 0.600 & 0.000 & 1.000 \\
5  & 2  & 17 & 3 - 5 years         & I'm not sure     & 0.000 & 0.562 & large  & T & 1.000 & 0.000 & 0.053 & 0.947 \\
6  & 3  & 24 & 6 - 10 years        & Not sure         & 0.000 & 0.322 & medium & T & 0.111 & 0.889 & 0.000 & 1.000 \\
7  & 3  & 18 & More than 10        & Yes              & 0.001 & 0.552 & large  & T & 0.938 & 0.062 & 0.391 & 0.609 \\
8  & 2  & 15 & More than 10        & Yes              & 0.001 & 0.316 & medium & T & 0.469 & 0.531 & 0.170 & 0.830 \\
9  & 3  & 23 & 1 - 2 years         & None             & 0.001 & 0.545 & large  & T & 0.333 & 0.667 & 0.000 & 1.000 \\
10 & 1  & 24 & Manager             & Not sure         & 0.001 & 0.304 & medium & F & 0.100 & 0.900 & 0.000 & 1.000 \\
11 & 3  & 15 & More than 10        & Yes              & 0.001 & 0.301 & medium & F & 0.593 & 0.407 & 0.256 & 0.744 \\
12 & 3  & 17 & Less than one year  & SaaS             & 0.002 & 0.497 & medium & F & 1.000 & 0.000 & 0.135 & 0.865 \\
13 & 2  & 23 & 1 - 2 years         & Hardware roaming & 0.002 & 0.492 & medium & F & 1.000 & 0.000 & 0.194 & 0.806 \\
14 & 8  & 23 & I'm not sure        & I don't know     & 0.003 & 0.480 & medium & F & 1.000 & 0.000 & 0.079 & 0.921 \\
15 & 7  & 23 & Yes (2 countries)   & I don't know     & 0.003 & 0.480 & medium & F & 1.000 & 0.000 & 0.079 & 0.921 \\
16 & 4  & 18 & Yes                 & Yes              & 0.003 & 0.474 & medium & F & 0.818 & 0.182 & 0.353 & 0.647 \\
17 & 1  & 18 & Decision maker      & Yes              & 0.004 & 0.464 & medium & F & 1.000 & 0.000 & 0.483 & 0.517 \\
18 & 10 & 23 & No                  & None             & 0.004 & 0.458 & medium & F & 0.250 & 0.750 & 0.000 & 1.000 \\
19 & 5  & 17 & Technology          & SaaS             & 0.005 & 0.449 & medium & F & 0.455 & 0.545 & 0.071 & 0.929 \\
20 & 3  & 18 & 3 - 5 years         & Yes              & 0.006 & 0.441 & medium & F & 0.273 & 0.727 & 0.750 & 0.250 \\
21 & 1  & 23 & Architect           & None             & 0.008 & 0.424 & medium & F & 0.222 & 0.778 & 0.000 & 1.000 \\
22 & 10 & 23 & No                  & Platform         & 0.008 & 0.424 & medium & F & 0.500 & 0.500 & 0.097 & 0.903 \\
23 & 9  & 17 & Only cloud          & SaaS             & 0.009 & 0.420 & medium & F & 0.600 & 0.400 & 0.118 & 0.882 \\
24 & 3  & 17 & 1 - 2 years         & I'm not sure     & 0.010 & 0.410 & medium & F & 0.333 & 0.667 & 0.030 & 0.970 \\
25 & 1  & 23 & Manager             & Mobile app       & 0.011 & 0.405 & medium & F & 1.000 & 0.000 & 0.343 & 0.657 \\
26 & 6  & 17 & 250 - 999           & Both             & 0.014 & 0.393 & medium & F & 0.000 & 1.000 & 0.588 & 0.412 \\
27 & 3  & 18 & 1 - 2 years         & Yes              & 0.014 & 0.393 & medium & F & 0.167 & 0.833 & 0.697 & 0.303 \\
28 & 9  & 23 & Only cloud          & I don't know     & 0.019 & 0.376 & medium & F & 0.400 & 0.600 & 0.059 & 0.941 \\
29 & 5  & 16 & Health              & I'm not sure     & 0.021 & 0.369 & medium & F & 0.500 & 0.500 & 0.054 & 0.946 \\
30 & 1  & 17 & Security Engineer   & I'm not sure     & 0.021 & 0.369 & medium & F & 0.500 & 0.500 & 0.054 & 0.946 \\
31 & 7  & 18 & No (only 1 country) & Yes              & 0.022 & 0.368 & medium & F & 0.412 & 0.588 & 0.773 & 0.227 \\
32 & 9  & 17 & Yes - both          & SaaS             & 0.026 & 0.356 & medium & F & 0.121 & 0.879 & 0.500 & 0.500 \\
33 & 5  & 23 & Technology          & I don't know     & 0.028 & 0.352 & medium & F & 0.273 & 0.727 & 0.036 & 0.964 \\
34 & 9  & 23 & Only on-prem        & Platform         & 0.030 & 0.347 & medium & F & 1.000 & 0.000 & 0.158 & 0.842 \\
35 & 7  & 17 & Yes (2 countries)   & SaaS             & 0.030 & 0.347 & medium & F & 1.000 & 0.000 & 0.158 & 0.842 \\
36 & 7  & 16 & Yes (10 or more)    & I'm not sure     & 0.031 & 0.346 & medium & F & 0.188 & 0.812 & 0.000 & 1.000 \\
37 & 6  & 23 & below 50            & Hardware roaming & 0.035 & 0.337 & medium & F & 0.571 & 0.429 & 0.188 & 0.812 \\
38 & 6  & 17 & 250 - 999           & Software         & 0.036 & 0.336 & medium & F & 0.600 & 0.400 & 0.176 & 0.824 \\
39 & 4  & 17 & Yes                 & I'm not sure     & 0.040 & 0.328 & medium & F & 0.000 & 1.000 & 0.176 & 0.824 \\
40 & 10 & 17 & No                  & Software         & 0.043 & 0.325 & medium & F & 0.500 & 0.500 & 0.161 & 0.839 \\
41 & 9  & 23 & Yes - both          & I don't know     & 0.043 & 0.324 & medium & F & 0.061 & 0.939 & 0.333 & 0.667 \\
42 & 5  & 18 & Research \& edu.    & Yes              & 0.047 & 0.318 & medium & F & 0.333 & 0.667 & 0.700 & 0.300 \\
43 & 10 & 18 & I am not certain    & Yes              & 0.048 & 0.317 & medium & F & 0.286 & 0.714 & 0.688 & 0.312 \\
\hline
\end{tabular}
\captionsetup{justification=centering}
\caption{\small{Cramér’s V ($\chi_2$) test results for $p < 0.05$ and medium and large effects. \\ $^*$The (non-adaptive) one-stage linear step-up procedure (LSU) for controlling the false discovery rate.}}
\end{table}

\fi

\end{document}